\documentclass[prd,preprint,superscriptaddress,showpacs,byrevtex]{revtex4}
\usepackage{bm}
\def\fsl#1{\setbox0=\hbox{$#1$}           
   \dimen0=\wd0                                 
   \setbox1=\hbox{/} \dimen1=\wd1               
   \ifdim\dimen0>\dimen1                        
      \rlap{\hbox to \dimen0{\hfil/\hfil}}      
      #1                                        
   \else                                        
      \rlap{\hbox to \dimen1{\hfil$#1$\hfil}}   
      /                                         
   \fi}                                         %
\newcommand{\be}{\begin{equation}}
\newcommand{\ee}{\end{equation}}
\newcommand{\bea}{\begin{eqnarray}}
\newcommand{\eea}{\end{eqnarray}}
\newcommand{\beq}{\begin{equation}}
\newcommand{\eeq}{\end{equation}}
\newcommand{\beqs}{\begin{eqnarray}}
\newcommand{\eeqs}{\end{eqnarray}}

\begin{document}
\title{ Potential Energy Between Static Quarks Is Time Dependent in The Classical Yang-Mills Theory }
\author{Gouranga C Nayak }\thanks{G. C. Nayak was affiliated with C. N. Yang Institute for Theoretical Physics in 2004-2007.}
\affiliation{ C. N. Yang Institute for Theoretical Physics, Stony Brook University, Stony Brook NY, 11794-3840 USA}
\date{\today}
\begin{abstract}
Lattice QCD predicts that the potential energy between static quarks is independent of time. However, in this paper we show that the gauge invariant color singlet potential energy between static quarks in the classical Yang-Mills theory depends on time even if the quarks are at rest. This is a consequence of the time dependent fundamental color charge $q^a(t)$ of the quark in the classical Yang-Mills theory. We find that the gauge invariant color singlet time dependent potential energy between static quarks does not violate the conservation of energy in the Yang-Mills theory.
\end{abstract}
\maketitle
\pagestyle{plain}

\pagenumbering{arabic}

\section{Introduction}

Confinement is an unsolved problem in particle physics. The fundamental issue here is to know the exact form of the color potential produced by the color charge of the quark. Once we know the exact form of the color potential produced by the color charge of the quark then we can explain why the quarks are confined inside the hadron, {\it i. e.}, why we have not directly observed free quarks at the experiments.

After the Yang-Mills theory \cite{yang} was discovered in 1954 there has been much progress in quantum chromodynamcs (QCD). In the renormalized QCD \cite{th} the asymptotic freedom occurs \cite{gr,pol} where the renormalized coupling decreases at short distances and increases at long distances. Hence in the renormalized QCD the partonic scattering cross section at short distances can be calculated by using perturbative quantum chromodynamics (pQCD).

However, the formation of hadron from partons is a long distance phenomenon in QCD where the renormalized pQCD is not applicable. The formation of hadron from partons can be studied by using renormalized non-perturbative QCD. Note that the analytical solution of the renormalized non-pertubtaive QCD is not found yet. Because of this reason we depend on the numerical prediction from the lattice QCD. The numerical results from lattice QCD predicts that the color singlet potential energy between static quarks, which is the measure of the Wilson loop \cite{wl}, depends on the separation between quarks but is independent of time \cite{latall}. However, as we will show in this paper, we find that the gauge invariant color singlet potential energy between static quarks in the classical Yang-Mills theory depends on time even if the quarks are at rest. Note that as mentioned in section III of \cite{momsum} the potential energy in QCD at long distance is the same potential energy that is obtained in the classical Yang-Mills theory because the Yang-Mills theory was discovered by making analogy with the Maxwell theory by extending the U(1) gauge group to the SU(3) gauge group \cite{yang,nj,ne}.

Another limitation of lattice QCD is that it can not predict the form of the color potential $A_\nu^c(x)$ produced by the color charges of the quark where
$c=1,...,8$ are the color indices. Hence unlike electromagnetic potential $A^\nu(x)$ which has four components ($\nu =0,1,2,3$), the color potential $A_\nu^c(x)$ has 32 components. The lattice QCD predicts the color singlet potential energy $V(r)$ between static quarks but can not predict the color potential $A_\nu^c(x)$ produced by the quark.

It is worth mentioning here that although the Yang-Mills theory was discovered almost 65 years ago, the exact form of the color potential $A_\nu^c(x)$ produced by the quark has not been found yet although many attempts have been made in the past, see for example \cite{p1,p2,p3,p4}. The electrical (Coulomb) potential produced by the electric charge explains the atomic bound state such as hydrogen atom in the Bohr's atomic model or in the Schrodinger equation. Similarly the color potential $A_\nu^c(x)$ produced by the color charge of the quark can be used to study bound state hadron formation from quarks and/or antiquarks. Hence it is necessary to find the exact form of the color potential $A_\nu^c(x)$ produced by the quark.

Note that the fundamental scalar electric charge $e$ of the electron is constant in classical electrodynamics but the fundamental color charge ${\vec q}(t)$ of the quark is a time dependent eight dimensional vector in the color space in the adjoint representation of SU(3) in the classical Yang-Mills theory \cite{ne}. The general form of the fundamental time dependent color charge vector ${\vec q}(t)$ of the quark in the classical Yang-Mills theory is given in \cite{ne}. At this stage it is necessary to point out that in the Yang-Mills theory the quark field $\psi_j(x)$ has three color indices $j=1,2,3$=RED, BLUE, GREEN which implies that RED, BLUE, GREEN symbols are not color charges of the quark but ${\vec q}(t)$ is the fundamental time dependent color charge of the quark \cite{ne}.

The color potential $A_\nu^{c}(x)$ produced by the color charge ${\vec q}(t)$ of the quark in motion is given by \cite{nj}
\bea
A_\nu^c(x) = \frac{u_\nu(\tau_0)}{\tau'}q^d(\tau_0)[\frac{{\rm exp}[g\int d\tau' \frac{C(\tau_0)}{\tau'}]-1}{g\int d\tau' \frac{C(\tau_0)}{\tau'}}]_{cd}
\label{1cpx}
\eea
where the indefinite integral is $d\tau'$ and $u^\nu(\tau)=\frac{ds^\nu(\tau)}{d\tau}$ is the four-velocity of the quark and
\bea
C^{bd}(t) = f^{bda}q^a(t),~~~~~~~~~~\tau' = u(\tau_0) \cdot (x-s(\tau_0)),~~~~~~~~~~~~~~x_0-s_0(\tau_0)=|{\vec x}-{\vec s}(\tau_0)|.
\label{2cpx}
\eea
It is important to remember that for the constant color charge ${\vec q}$ the eq. (\ref{1cpx}) reduces to Maxwell-like (abelian-like) potential. This means when all the color charges $q^a$ are constants then the theory is an abelian-like theory [like U(1) gauge theory] but not the SU(3) gauge theory. Hence all the color charges $q^a$ are not constants in the classical Yang-Mills theory. This implies that there is no gauge choice in the classical Yang-Mills theory where all the color charges $q^a$ are constants, see sections VII, VIII and IX for more details.

Due to the time dependent fundamental color charge ${\vec q}(t)$ of the quark, many gauge invariant quantities in the classical Yang-Mills theory are time dependent even if the quarks are at rest. We consider such a gauge invariant quantity in this paper which is the color singlet potential energy between static quarks in the classical Yang-Mills theory.

We find in this paper we find that the gauge invariant color singlet potential energy between static quarks in the classical Yang-Mills theory depends on time even if the quarks are at rest. We find that the gauge invariant color singlet time dependent potential energy between static quarks does not violate the conservation of energy in the Yang-Mills theory.

The paper is organized as follows. In section II we briefly review the derivation of potential energy between static charges from the electric field energy. In section III we derive the general form of the time dependent chromo-electric field produced by the static quark in the Yang-Mills theory. In section IV we show that the gauge invariant color singlet potential energy between static quarks in the Yang-Mills theory is time dependent even if the quarks are at rest. In section V we show that the gauge invariant color singlet time dependent potential energy between static quarks does not violate the conservation of energy in the Yang-Mills theory. In section VI
we discuss the consistency with the energy conservation equation from the gauge invariant Noether's theorem in the Yang-Mills theory. In section VII we discuss time dependent gauge transformation of the time dependent fundamental color charge of the quark. In section VIII we show that the constant color charge ${\vec q}$ produces coulomb potential. In section IX we show that the time dependent fundamental color charge of the quark in classical Yang-Mills theory is not gauge transformation of constant color charge. In section X we show that the gauge invariant color singlet time dependent potential energy between static quarks in the Yang-Mills theory is not a gauge artifact. Section XI contains conclusions.

\section{Potential Energy Between Static Charges From Electric Field Energy }\label{ese}

Recall that the form of the potential energy between static electrons in the classical electrodynamics can be derived from the electric field energy produced by the static electrons. Similarly we find in this paper that the potential energy between static quarks in the classical Yang-Mills theory can be obtained from the chromo-electric field energy produced by the static quarks.

For this reason we will first briefly review the derivation of the potential energy between static electrons in the classical electrodynamics from the electric field energy produced by the static electrons in this section before discussing the corresponding situation in the Yang-Mills theory in the remaining of the paper. Although the derivation of the potential energy between static electrons in the classical electrodynamics from the electric field energy produced by the static electrons is well known in the literature but we will briefly review its derivation in this section because we will use the similar technique to study the potential energy between static quarks from the chromo-electric field energy produced by the static quarks in the Yang-Mills theory in this paper.

In Maxwell theory the Coulomb electric field produced by a static electron of charge $e$ is given by $e\frac{{\hat r}}{r^2}$. For a system of static electrons the total electric field energy is given by
\bea
E_f= \frac{1}{2} \int d^3r {\vec E}(r) \cdot {\vec E}(r)
\label{ef2ac}
\eea
which is independent of time where ${\vec E}(r)$ is the electric field produced by the static electrons. In order to derive the potential energy between two static electrons from the electric field energy from eq. (\ref{ef2ac}), let us consider one electron at the origin and another electron at a distance $R$ along the z-axis. The electric field at any position ${\vec r}$ is given by
\bea
{\vec E}(r) = e\frac{{\hat r}}{r^2} + e\frac{{\vec r}-{\hat z}R}{|{\vec r}-{\hat z}R|^3}.
\label{e1}
\eea
By using eq. (\ref{e1}) in (\ref{ef2ac}) we find that the total electric field energy is given by
\bea
&& E_f= e^2\frac{1}{2} \int d^3r \frac{1}{r^4} + e^2\frac{1}{2} \int d^3r \frac{1}{|{\vec r}-{\hat z}R|^4} +e^2 \int d^3r \frac{{\hat r} \cdot ({\vec r}-{\hat z}R)}{r^2|{\vec r}-{\hat z}R|^3}=E_{f}^{11}+E_{f}^{22}+E_{f}^{12}\nonumber \\
\label{e2}
\eea
where
\bea
E_{f}^{11}=e^2\frac{1}{2} \int d^3r \frac{1}{r^4} = \infty,~~~~~~~~E_{f}^{22}= e^2\frac{1}{2} \int d^3r \frac{1}{|{\vec r}-{\hat z}R|^4} =\infty
\label{e3}
\eea
are the infinite self energies of the electrons and
\bea
&& \frac{E_{f}^{12}}{4\pi}=V(R) =\frac{e^2}{4\pi} \int d^3r \frac{{\hat r} \cdot ({\vec r}-{\hat z}R)}{r^2|{\vec r}-{\hat z}R|^3}= \frac{e^2}{2} \int_0^\infty  dr  \frac{1}{r^2}[1+\frac{r-R}{|r- R|}]  =  \frac{e^2}{R}
\label{e4}
\eea
is the finite potential energy between two static electrons separated by distance $R$.

For static electrons we have ${\vec E}(r)=-{\vec \nabla}A_0(r)$ which means the eq. (\ref{ef2ac}) can be written as
\bea
&& E_f = \frac{1}{2} \int dS {\hat n} \cdot [A_0(r) {\vec \nabla} A_0(r)]+\frac{1}{2} \int d^3r A_0(r) \rho(r)
\label{ef4p}
\eea
where $\rho(r)=j_0(r)$ is the charge density, $S$ is the surface enclosing the volume and ${\hat n}$ is the unit normal to the surface. Since the boundary surface term at infinity vanishes, {\it i. e.},
\bea
\frac{1}{2} \int dS {\hat n} \cdot [A_0(r) {\vec \nabla} A_0(r)] =0
\label{ef4q}
\eea
we find from eq. (\ref{ef4p})
\bea
E_f=\frac{1}{2} \int d^3r A_0(r) \rho(r).
\label{ef6}
\eea
Using the charge density for a point electron at rest
\bea
\rho(r) =e \delta^{(3)}({\vec r}-{\vec r}_i)
\label{ef9}
\eea
and neglecting the self energies we reproduce eq. (\ref{e4}) from eq. (\ref{ef6}).

Hence from eq. (\ref{e4}) we find that the potential energy between static electrons can be obtained from the electric field energy $\frac{1}{2} \int d^3r {\vec E}(r)\cdot {\vec E}(r)$ produced by the static electrons in Maxwell theory. Similarly the potential energy between static quarks can be obtained from the chromo-electric field energy $\frac{1}{2} \int d^3r {\vec E}^c(t,r)\cdot {\vec E}^c(t,r)$ produced by the static quarks in the Yang-Mills theory which we will discuss in this paper.

\section{ Time Dependent Chromo-Electric Field Produced By Static Quark }\label{qu}

From eq. (\ref{1cpx}) we find that the color potential produced by the color charge ${\vec q}(t)$ of the quark at rest in the Yang-Mills theory is given by
\bea
A_0^c(t,r) = \frac{q^d(t-r)}{r}[\frac{{\rm exp}[g\int dr \frac{C(t-r)}{r}]-1}{g\int dr \frac{C(t-r)}{r}}]_{cd}
\label{3cpx}
\eea
which depends on time $t$ even if the quark is at rest where $C^{ab}(t)$ is given by eq. (\ref{2cpx}) and
\bea
{\vec A}^c(t,r) =0.
\label{4cpx}
\eea
We have used the natural unit in this paper.

In the Yang-Mills theory the chromo-electric field ${\vec E}^c(t,r)$ and the chromo-magnetic field ${\vec B}^c(t,r)$ are given by
\bea
E^{kc}(t,r) = F^{k0 c}(t,r),~~~~~~~~~~~~B^{kc}(t,r)=-\frac{1}{2} \epsilon^{klm} F^{lm c}(t,r)
\label{5cpx}
\eea
where
\bea
F_{ \lambda \delta}^c(t,r) = \partial_\lambda A_\delta^c(t,r) -\partial_\delta A_\lambda^c(t,r)+gf^{cdh} A_\lambda^d(t,r)A_\delta^h(t,r).
\label{6cpx}
\eea
From eqs. (\ref{5cpx}), (\ref{6cpx}), (\ref{4cpx}) and (\ref{3cpx}) we find that the chromo-electric field ${\vec E}^c(t,r)$ produced by the color charge ${\vec q}(t)$ of the quark at rest is given by
\bea
{\vec E}^c(t,r) =-{\vec \nabla}\left[ \frac{q^d(t-r)}{r}[\frac{{\rm exp}[g\int dr \frac{C(t-r)}{r}]-1}{g\int dr \frac{C(t-r)}{r}}]_{cd}\right]
\label{7cpx}
\eea
which depends on time $t$ even if the quark is at rest.

Simplifying the infinite number of non-communing terms we find in the adjoint representation of SU(3) \cite{nj}
\bea
\partial_\nu [e^{gK(x)}-1]_{cd}=[\partial_\nu \omega^b(x)] [\frac{e^{gK(x)}-1}{gK(x)}]_{hb}gf^{hce}[e^{gK(x)}]_{ed},~~~~~~~~K_{cd}(x)=f^{cdh}\omega^h(x).
\label{gn3}
\eea
From
\bea
[gK(x)]_{cd}[\frac{1}{gK(x)}]_{db}=\delta_{cb}
\eea
we find
\bea
\partial_\nu [\frac{1}{gK(x)}]_{ab}=-[\frac{1}{gK(x)}]_{ap}gf^{pcd}\partial_\nu \omega^d(x)[\frac{1}{gK(x)}]_{cb}.
\label{gn5s}
\eea
Using
\bea
&&\partial_\nu [\frac{e^{gK(x)}-1}{gK(x)}]_{cd}=\partial_\nu [[e^{gK(x)}-1]_{ca}[\frac{1}{gK(x)}]_{ad}]\nonumber \\
&&=[\partial_\nu [e^{gK(x)}-1]_{ca}][\frac{1}{gK(x)}]_{ad}+ [e^{gK(x)}-1]_{ca}[\partial_\nu [\frac{1}{gK(x)}]_{ad}]
\label{gn4}
\eea
we find from eqs. (\ref{gn4}), (\ref{gn3}) and (\ref{gn5s}) that
\bea
&&\partial_\nu [\frac{e^{gK(x)}-1}{gK(x)}]_{cd}=[\partial_\nu \omega^h(x)] [\frac{e^{gK(x)}-1}{gK(x)}]_{ph}gf^{pcs}[\frac{e^{gK(x)}}{gK(x)}]_{sd}- [\frac{e^{gK(x)}-1}{gK(x)}]_{cp}gf^{psh}\partial_\nu \omega^h(x)[\frac{1}{gK(x)}]_{sd}.\nonumber \\
\label{fna}
\eea
Eq. (\ref{fna}) can also be obtained if we use
\bea
&&\partial_\nu [\frac{e^{gK(x)}-1}{gK(x)}]_{cd}=\partial_\nu [[\frac{1}{gK(x)}]_{cb}[e^{gK(x)}-1]_{bd}]\nonumber \\
&&=[\partial_\nu [\frac{1}{gK(x)}]_{cb}][e^{gK(x)}-1]_{bd}+[\frac{1}{gK(x)}]_{cb}[\partial_\nu [e^{gK(x)}-1]_{bd}]
\label{gn4s}
\eea
instead of using eq. (\ref{gn4}) as it should be the case.

Hence from eq. (\ref{fna}) we find
\bea
&&\frac{d}{dr}[\frac{{\rm exp}[g\int dr \frac{C(t-r)}{r}]-1}{g\int dr \frac{C(t-r)}{r}}]_{bd}=[\frac{q^a(t-r)}{r}] [\frac{e^{g\int dr \frac{C(t-r)}{r}}-1}{g\int dr \frac{C(t-r)}{r}}]_{pa}gf^{pbs}[\frac{e^{g\int dr \frac{C(t-r)}{r}}}{g\int dr \frac{C(t-r)}{r}}]_{sd}\nonumber \\
&&- [\frac{e^{g\int dr \frac{C(t-r)}{r}}-1}{g\int dr \frac{C(t-r)}{r}}]_{bp}gf^{psa}[\frac{q^a(t-r)}{r}][\frac{1}{g\int dr \frac{C(t-r)}{r}}]_{sd}.
\label{fnb}
\eea
From eqs. (\ref{7cpx}) and (\ref{fnb}) we find that the chromo-electric field $ {\vec E}^c(t,r)$ produced by the color charge ${\vec q}(t)$ of the quark at rest in the Yang-Mills theory is given by
\bea
&&{\vec E}^b(t,r) ={\hat r} \frac{q^a(t-r)}{r^2}[\frac{{\rm exp}[g\int dr \frac{C(t-r)}{r}]-1}{g\int dr \frac{C(t-r)}{r}}]_{ba}
-\frac{{\hat r}}{r} \frac{dq^a(t-r)}{dr}[\frac{{\rm exp}[g\int dr \frac{C(t-r)}{r}]-1}{g\int dr \frac{C(t-r)}{r}}]_{ba} \nonumber \\
&&-{\hat r}\{ [\frac{e^{g\int dr \frac{C(t-r)}{r}}-1}{g\int dr \frac{C(t-r)}{r}}]_{pd}gf^{pbs}- [\frac{e^{g\int dr \frac{C(t-r)}{r}}-1}{g\int dr \frac{C(t-r)}{r}}]_{bp}gf^{psd}\} [\frac{1}{g\int dr \frac{C(t-r)}{r}}]_{sa}\frac{q^a(t-r)q^d(t-r)}{r^2}\nonumber \\
\label{fnc}
\eea
which depends on time $t$ even if the quark is at rest where $C^{ab}(t)$ is given by eq. (\ref{2cpx}).

Note that when the color charge ${\vec q}$ is constant we find from eq. (\ref{fnc})
\bea
&&{\vec E}^c(r) ={\hat r} \frac{q^c}{r^2}
\label{fncs}
\eea
which is the Coulomb-like field similar to Maxwell theory. This means when all the color charges $q^a$ are constants then the theory is an abelian-like theory [like U(1) gauge theory] but not the SU(3) gauge theory. Hence all the color charges $q^a$ are not constants in the classical Yang-Mills theory. This implies that there is no gauge choice in the classical Yang-Mills theory where all the color charges $q^a$ are constants, see sections VII, VIII and IX for more details.

\section{ Time Dependent Potential Energy Between Static Quarks in Classical Yang-Mills Theory }\label{qu3}

From eq. (\ref{ef6}) we find that the potential energy between static electrons in classical electrodynamics is given by
\bea
V=\frac{1}{8\pi} \int d^3r A_0(r) \rho(r).
\label{pote}
\eea
However, the potential energy between static quarks in the classical Yang-Mills theory is not given by $\frac{1}{8\pi} \int d^3r A_0^c(t,r) \rho^c(t,r)$ if the boundary surface term does not vanish, see eq. (\ref{nvst}), where the Yang-Mills color charge density $\rho_0(t,r)$ of the quarks is given by
\bea
\rho^c(t,r)=j_0^c(t,r),~~~~~~~~~~~j_\lambda^{~c}(x)=D^\delta[A]F_{\delta \lambda}^c(x),~~~~~~~~~~~~~~~~~~~~~D_\lambda^{cd}[A]=\delta^{cd}\partial_\lambda +gf^{cad}A_\lambda^a(x).\nonumber \\
\label{ceq}
\eea
The Yang-Mills field tensor $F_{\lambda \delta}^c(x)$ in eq. (\ref{ceq}) is given by eq. (\ref{6cpx}).

First of all note that, as mentioned earlier, there can be two types of fundamental charges in the nature: 1) constant electric charge $e$ of the electron [U(1) gauge theory] and 2) time dependent color charge ${\vec q}(t)$ of the quark [SU(3) gauge theory]. As mentioned earlier the color charge ${\vec q}^a(t)$ of the quark can not be constant because for constant color charge ${\vec q}$ we find from eq. (\ref{fncs})
\bea
{\vec E}^c(r)\cdot {\vec E}^c(r)=\frac{q^cq^c}{r^4}
\label{ccce}
\eea
which has the Coulomb form similar to abelian-like theory [like U(1) gauge theory]. It should be mentioned here that, similar to abelian electric charge density $\rho(r)=e\delta^{(3)}({\vec r}-{\vec r}_i)$ of static electron in eq. (\ref{ef9}) in Maxwell theory, an abelian-like color charge density
\bea
{\cal J}_0^c(x)=gT^c\delta^{(3)}({\vec r}-{\vec r}_i)
\label{ablcd}
\eea
which has been used in the literature for static quark, see for example \cite{fisc}, is not correct because ${\cal J}_0^c(x)$ is a vector in color space whereas $T^c$ is a matrix in color space [Gell-Mann matrix] with components $T^c_{ij}$. In addition to this the color charge density in eq. (\ref{ablcd}) is not consistent with the Yang-Mills theory because the Yang-Mills color charge density $j_0^c(t,r)=\rho^c(t,r)$ of the static quark is time dependent even if the quark is at rest and the Yang-Mills color charge density $j_0^c(t,r)=\rho^c(t,r)$ of the static quark contains infinite powers of $g$, see \cite{nj,ne} for details where we have shown that for static quark in Yang-Mills theory
\bea
j_0^c(t,r)=\rho^c(t,r)\neq gT^c\delta^{(3)}({\vec r}-{\vec r}_i).
\label{nablcd}
\eea

For quarks at rest we find from eqs. (\ref{3cpx}), (\ref{fnc}), (\ref{4cpx}), (\ref{5cpx}) and (\ref{6cpx}) that the chromo-electric field produced by the static quarks is given by
\bea
{\vec E}^c(t,r) = -{\vec \nabla} A_0^c(t,r)
\label{yef3}
\eea
which is time dependent even if the quarks are at rest and
\bea
{\vec B}^c(t,r) =0
\label{8cpx}
\eea
where $A_0^c(t,r)$ is the zero component of the color potential $A_\nu^c(t,r)$ produced by the quarks. Hence for static quarks the (color) field energy is the chromo-electric field energy given by
\bea
E_f(t)=\frac{1}{2} \int d^3r~{\vec E}^c(t,r) \cdot {\vec E}^c(t,r)
\label{9cpx}
\eea
which depends on time $t$ even if the quarks are at rest, {\it i. e.} for static quarks we have
\bea
\frac{dE_f(t)}{dt}=\frac{d[\frac{1}{2} \int d^3r~{\vec E}^c(t,r) \cdot {\vec E}^c(t,r)]}{dt} \neq 0
\label{11cpxa}
\eea
because the chromo-electric field produced by the static quark depends on time, see eq. (\ref{fnc}).

Using eq. (\ref{yef3}) in (\ref{9cpx}) we find
\bea
&&E_f(t)=\frac{1}{2} \int dS {\vec n} \cdot [A_0^c(t,r) {\vec \nabla} A_0^c(t,r)]-\frac{1}{2} \int d^3r A_0^c(t,r) {\nabla}^2 A_0^c(t,r)
\label{yef4p}
\eea
where we have used the divergence theorem. From eqs. (\ref{ceq}), (\ref{yef3}),  (\ref{8cpx}) and (\ref{4cpx}) we have for static quarks
\bea
\rho^c(t,r) = {\vec \nabla} \cdot {\vec E}^c(t,r) = -{\nabla}^2 A_0^c(t,r)
\label{abf3}
\eea
which when used in eq. (\ref{yef4p}) gives
\bea
&&E_f(t)=\frac{1}{2} \int d^3r {\vec E}^c(t,r) \cdot {\vec E}^c(t,r)=\frac{1}{2} \int dS {\vec n} \cdot [A_0^c(t,r) {\vec \nabla} A_0^c(t,r)]+\frac{1}{2} \int d^3r A_0^c(t,r) \rho^c(t,r)\nonumber \\
\label{abf4}
\eea
which is similar to eq. (\ref{ef4p}) in Maxwell theory.

However, unlike the vanishing boundary surface term in Maxwell theory in eq. (\ref{ef4q}), the boundary surface term $\frac{1}{2} \int dS {\vec n} \cdot [A_0^a(t,r) {\vec \nabla} A_0^a(t,r)]$ in eq. (\ref{abf4}) for static quarks does not vanish in the SU(3) classical Yang-Mills theory. This is due to the following reason.

The electric field energy density $\frac{1}{2} {\vec E}(r) \cdot {\vec E}(r)$ produced by static electrons falls off as $\frac{1}{r^4}$ which from the electric field energy $\frac{1}{2}\int d^3r {\vec E}(r) \cdot {\vec E}(r)$ produced by static electrons predicts that the potential energy $V(R)$ between two static electrons separated by distance $R$ falls of as $\frac{1}{R}$, see eq. (\ref{e4}). Note that confinement happens in QCD at long distance. As mentioned in section III of \cite{momsum}, the potential energy at infinite distance in QCD and the potential energy at infinite distance in the classical Yang-Mills theory are same because the Yang-Mills theory was discovered by making analogy with the Maxwell theory by extending the U(1) gauge group to the SU(3) gauge group \cite{yang,nj,ne}. Hence due to the confinement in QCD we find from the gauge invariant chromo-electric field energy $\frac{1}{2} \int d^3r {\vec E}^c(t,r) \cdot {\vec E}^c(t,r)$ produced by the static quarks that the gauge invariant chromo-electric field energy density $\frac{1}{2}{\vec E}^c(t,r) \cdot {\vec E}^c(t,r)$ at infinite distance produced by the static quarks in the classical Yang-Mills theory does not fall off faster than $\frac{1}{r^3}$. This implies that that $[{\vec \nabla} A_0^c(t,r)]^2$ at infinite distance produced by static quarks does not fall of faster than $\frac{1}{r^3}$ which means the color potential $A_0^c(t,r)$ at infinite distance produced by the static quarks does not fall off faster than $\frac{1}{r^{\frac{1}{2}}}$ and the ${\vec \nabla} A_0^c(t,r)$ at infinite distance produced by static quarks does not fall off faster than $\frac{1}{r^{\frac{3}{2}}}$. Hence we find that $A_0^c(t,r) {\vec \nabla} A_0^c(t,r)$ at infinite distance produced by static quarks in the classical Yang-Mills theory does not fall off faster than $\frac{1}{r^2}$ which implies that
\bea
\frac{1}{2} \int dS {\vec n} \cdot [A_0^c(t,r) {\vec \nabla} A_0^c(t,r)]\neq 0
\label{nvst}
\eea
when the boundary surface is at the infinite distance. The boundary surface term in eq. (\ref{nvst}) is also non-zero when the boundary surface is at finite distance.

Since $\int dS {\vec n} \cdot [A_0^c(t,r) {\vec \nabla} A_0^c(t,r)]$ is non-zero and is not gauge invariant we find from eqs. (\ref{nvst}) and (\ref{abf4}) that $\int d^3r A_0^c(t,r) \rho^c(t,r)$ is not gauge invariant. This implies that, unlike the potential energy $\frac{1}{8\pi} \int d^3r A_0(r) \rho(r)$ in eq. (\ref{pote}) for static electrons in Maxwell theory, the $\frac{1}{8\pi} \int d^3r A_0^c(t,r) \rho^c(t,r)$ is not the potential energy between static quarks in Yang-Mills theory because it is not gauge invariant. When the gauge non-invariant non-vanishing boundary surface term $\frac{1}{2} \int dS {\vec n} \cdot [A_0^c(t,r) {\vec \nabla} A_0^c(t,r)]$ is added to the gauge non-invariant $\frac{1}{2} \int d^3r A_0^c(t,r) \rho^c(t,r)$ then we obtain the gauge invariant chromo-electric field energy $\frac{1}{2} \int d^3r {\vec E}^c(t,r) \cdot {\vec E}^c(t,r)$ in eq. (\ref{abf4}). This implies that the gauge invariant potential energy between static quarks is obtained from the gauge invariant chromo-electric field energy $\frac{1}{2} \int d^3r {\vec E}^c(t,r) \cdot {\vec E}^c(t,r)$ produced by the static quarks.

Note from eqs. (\ref{pote}), (\ref{ef4p}), (\ref{ef4q}), (\ref{ef6}) and (\ref{ef2ac}) that the potential energy $V$ between static electrons in Maxwell theory is given by
\bea
V= \frac{1}{8\pi} \int d^3r {\vec E}(r) \cdot {\vec E}(r).
\label{potef}
\eea
Similarly from eqs. (\ref{abf4}) and (\ref{nvst}) we find that the color singlet gauge invariant potential energy $V(t)$ between static quarks in the classical Yang-Mills theory is given by
\bea
V(t)= \frac{1}{8\pi} \int d^3r {\vec E}^c(t,r) \cdot {\vec E}^c(t,r)
\label{ypotef}
\eea
which depends on time $t$ even if the quarks are at rest, see eq. (\ref{fnc}).

Hence we find that the color singlet gauge invariant potential energy $V(t)$ between static quarks in the classical Yang-Mills theory is given by eq. (\ref{ypotef})
which depends on time $t$ even if the quarks are at rest.

\section{ Time Dependent Potential Energy Between Static Quarks Does Not Violate Conservation of Energy in Yang-Mills Theory }\label{qu1}

In the previous section we saw that the color singlet gauge invariant potential energy $V(t)$ between static quarks in the Yang-Mills theory depends on time even if the quarks are at rest. This, however, does not violate the conservation of energy in the Yang-Mills theory which can be seen as follows.

For static quarks the chromo-magnetic field ${\vec B}^c(t,r)$ is zero, see eq. (\ref{8cpx}). From the Yang-Mills equation we have
\bea
D^\delta[A]F_{\delta \lambda}^c(x)=g{\bar \psi}(x)T^c\gamma_\lambda \psi(x)
\label{dfj}
\eea
where $\psi_k(x)$ is the Dirac field of the quark with $k=1,2,3$ being the color index (in this paper the suppression of color index $k$ in the quark field $\psi_k(x)$ is understood).

\subsection{ Energy Conservation in Source (Quark) Free Volume }

For $\lambda=k$ we find from eqs. (\ref{4cpx}) and (\ref{dfj}) that the chromo-electric field ${\vec E}^c(t,r)$ produced by static quarks in the source (quark) free region satisfies the equation
\bea
\frac{\partial {\vec E}^c(t,r)}{\partial t} =-gf^{cad} A_0^a(t,r) {\vec E}^d(t,r)
\label{10cpx}
\eea
which implies that the gauge invariant chromo-electric field energy produced by the static quarks in the source (quark) free volume $V$ is independent of time, {\it i. e.},
\bea
\frac{1}{2} \int dV~{\vec E}^c(t,r) \cdot \frac{\partial {\vec E}^c(t,r)}{\partial t}=0.
\label{t9cpx}
\eea
Note that eq. (\ref{t9cpx}) does not violate eq. (\ref{11cpxa}) because eq. (\ref{t9cpx}) is valid for source (quark) free volume $V$ whereas the volume integral $\int d^3r$ in eq. (\ref{11cpxa}) includes the sources (static quarks), see eq. (\ref{11xcp}).

\subsection{ Energy Conservation in The Volume Containing Sources (Static Quarks) }

For $\lambda=k$ we find from eqs. (\ref{4cpx}) and (\ref{dfj}) that the chromo-electric field ${\vec E}^a(t,r)$ produced by static quarks in the region containing sources (static quarks) satisfies the equation
\bea
\frac{\partial {\vec E}^c(t,r)}{\partial t} =-gf^{cad} A_0^a(t,r) {\vec E}^d(t,r)-g{\bar \psi}(x)T^c{\vec \gamma} \psi(x)
\label{10xcp}
\eea
which gives
\bea
\frac{dE_f(t)}{dt}=\frac{1}{2} \int d^3r~{\vec E}^c(t,r) \cdot \frac{\partial {\vec E}^c(t,r)}{\partial t}=-\int d^3r g{\bar \psi}(t,r)T^c {\vec \gamma} \psi(t,r) \cdot {\vec E}^a(t,r)
\label{11xcp}
\eea
which is the energy conservation equation for static quarks in the Yang-Mills theory.

Note that, unlike static electrons in classical electrodynamics where ${\vec j}(x)=0$, the static quarks in classical Yang-Mills theory gives the non-zero vector component of the Yang-Mills color current ${\vec j}^c(x) $ which can be seen from eqs. (\ref{3cpx}), (\ref{yef3}) and (\ref{10xcp}) which gives \cite{ne}
\bea
{\vec j}^c(x) =g{\bar \psi}(t,r)T^c {\vec \gamma} \psi(t,r)\neq 0,~~~~~~~~~~~~~~{\rm for~~static~~quarks}.
\label{ccu}
\eea
From eq. (\ref{ccu}) we find that the right hand side of eq. (\ref{11xcp}) is non-zero for static quarks which means even if the rate of work done on static electrons in Maxwell theory is zero but the rate of work done on static quarks in the Yang-Mills theory is non-zero.

Eq. (\ref{11xcp}) is the Poynting's theorem for static quarks in the Yang-Mills theory which can also be derived from the gauge invariant Noether's theorem in Yang-Mills theory, see eq. (\ref{yct8}).

Hence from eqs. (\ref{9cpx}), (\ref{ypotef}) and (\ref{11xcp}) we find that the gauge invariant color singlet time dependent potential energy between static quarks does not violate the conservation of energy in the Yang-Mills theory.

\section{ Consistency With The Energy Conservation Equation From The Gauge Invariant Noether's Theorem in Yang-Mills Theory}
 \label{gnoety}

The gauge invariant Noether's theorem in Dirac-Maxwell theory is described in \cite{nkgint} and the gauge invariant Noether's theorem in Yang-Mills theory is described in \cite{nkginty}. From the Dirac equation of $\psi(x)$ and ${\bar \psi}(x)$ of the quark in the presence of Yang-Mills potential $A_\nu^c(x)$ \cite{nkginty} and by using the properties of the Dirac matrices we find
\bea
&&\partial_\delta[\frac{i}{4} {\bar \psi}(x)[\gamma^\delta  ({\overrightarrow \partial}^\lambda -igT^cA^{\lambda c}(x)) +\gamma^\lambda  ({\overrightarrow \partial}^\delta -igT^cA^{\mu c}(x))\nonumber \\
&&-\gamma^\delta ({\overleftarrow \partial}^\lambda +igT^cA^{\lambda c}(x)) -\gamma^\lambda ({\overleftarrow \partial}^\delta +igT^cA^{\delta c}(x))  ] \psi(x)]\nonumber \\
&&=\partial_\delta [\frac{i}{2} {\bar \psi}(x)[\gamma^\delta  ({\overrightarrow \partial}^\lambda -igT^cA^{\lambda c}(x)) -\gamma^\delta ({\overleftarrow \partial}^\lambda +igT^cA^{\nu c}(x)) ] \psi(x)]\nonumber \\
&&=-g{\bar \psi}(x) T^c\gamma_\delta \psi(x) F^{\delta \lambda c}(x)
\label{yct4}
\eea
where $F_{\delta \lambda }^c(x)$ is given by eq. (\ref{6cpx}). Using eq. (\ref{yct4}) in the gauge invariant Noether's theorem in Yang-Mills theory \cite{nkginty} we find
\bea
\partial_\delta [F^{\delta \sigma c}(x) F_\sigma^{~~\lambda c}(x) +\frac{1}{4}g^{\delta \lambda}  F_{\sigma \nu}^c(x)F^{\sigma \nu c}(x)]
=g{\bar \psi}(x)T^c \gamma_\delta \psi(x) F^{\delta \lambda c}(x).
\label{yct7}
\eea
For $\lambda=0$ we find from eq. (\ref{yct7})
\bea
\frac{d}{dt}\int d^3x \frac{{\vec E}^c(x) \cdot {\vec E}^c(x)+{\vec B}^c(x) \cdot {\vec B}^c(x)}{2}=-\int d^3x {\vec \nabla}\cdot [{\vec E}^c(x) \times {\vec B}^c(x)]-\int d^3x g{\bar \psi}(x)T^c {\vec \gamma} \psi(x) \cdot {\vec E}^c(x)\nonumber \\
\label{yct8}
\eea
which is the energy conservation equation in the Yang-Mills theory.

Eq. (\ref{yct8}) is the Poynting theorem in the Yang-Mills theory.

For static quarks the chromo-magnetic field is zero, see eq. (\ref{8cpx}). Hence for static quarks in the Yang-Mills theory we find from eq. (\ref{yct8}) the energy conservation equation
\bea
\frac{1}{2} \int d^3r~{\vec E}^c(t,r) \cdot \frac{\partial {\vec E}^c(t,r)}{\partial t}=-\int d^3r g{\bar \psi}(t,r)T^c {\vec \gamma} \psi(t,r) \cdot {\vec E}^a(t,r)
\label{11xcpf}
\eea
which reproduces eq. (\ref{11xcp}) which was directly obtained by using the Yang-Mills equation. As shown in eq. (\ref{ccu}) the right hand side of eq. (\ref{11xcpf}) is non-zero for static quarks which means even if the rate of work done on static electrons in Maxwell theory is zero but the rate of work done on static quarks in the Yang-Mills theory is non-zero.

Hence from eqs. (\ref{9cpx}), (\ref{ypotef}), (\ref{11xcp}) and (\ref{11xcpf}) we find that the gauge invariant color singlet time dependent potential energy between static quarks does not violate the conservation of energy in the Yang-Mills theory.

\section{ Time Dependent Gauge Transformation of The Time Dependent Fundamental Color Charge of The Quark }\label{gtcc}

Under gauge transformations, the time dependent fundamental color charge $q^a(t)$ of the quark transforms homogeneously under the adjoint representation of the gauge group \cite{ne}
\bea
q'^a(t)=S_{ab}(t)q^b(t)=[e^{M(t)}]_{ab}q^b(t),~~~~~~~~~~~~~M_{ab}(t)=f^{abc}\beta^c(t)
\label{s3}
\eea
in SU(3) local gauge theory (the Yang-Mills theory in SU(3)) where $f^{abc}$ are the structure constants in SU(3) and $\beta^a(t)$ are the gauge transformation parameters. Similarly in SU(2) local gauge theory (in SU(2) Yang-Mills theory) one finds that the gauge transformed time dependent color charge $q_i(t)$ of a fermion is given by
\bea
q'_i(t)=S_{ij}(t)q_j(t)=[e^{M(t)}]_{ij}q_j(t),~~~~~~~~~~~~~M_{ij}(t)=\epsilon^{ijk}\beta^k(t)
\label{s2}
\eea
where $\epsilon^{ijk}$ are Levi-Civita symbol with $i,jk,k=1,2,3$.

Note that the three generators $G^i_{jk}$ of the SO(3) group are given by $G^i_{jk}=\epsilon^{ijk}$. Hence, as mentioned in \cite{ne}, the time dependent gauge transformation matrix $S(t)$ in the SU(2) Yang-Mills theory in eq. (\ref{s2}) is a time dependent rotation matrix $S(t)$ in SO(3). In terms of three time dependent Euler angles $ \beta'_1(t),~\beta'_2(t),~\beta'_3(t)$ the time dependent gauge transformation matrix $S(t)$ in eq. (\ref{s2}) in SU(2) Yang-Mills theory can be written as
\bea
S(t)=\left(\begin{array}{ccc}
{\rm cos}\beta'_1(t),&-{\rm sin}\beta'_1(t),&0\\
{\rm sin}\beta'_1(t),&{\rm cos}\beta'_1(t),&0\\
0,&0,&1\\
\end{array}\right)\left(\begin{array}{ccc}
{\rm cos}\beta'_2(t),&0,&{\rm sin}\beta'_2(t)\\
0,&1,&0\\
-{\rm sin}\beta'_2(t),&0,&{\rm cos}\beta'_2(t)\\
\end{array}\right)\left(\begin{array}{ccc}
{\rm cos}\beta'_3(t),&-{\rm sin}\beta'_3(t),&0\\
{\rm sin}\beta'_3(t),&{\rm cos}\beta'_3(t),&0\\
0,&0,&1\\
\end{array}\right).\nonumber \\
\label{rt}
\eea
For the discussion of the time dependent gauge transformation in the adjoint representation of SU(3) and the general rotation in SO(8), see sections 9 and 15 of \cite{ne}.

Note that the time dependent gauge transformation parameters $\beta^a(t)$ in the classical non-abelian Yang-Mills theory in eq. (\ref{s3}) are not arbitrary as they have to be consistent with the space-time dependent gauge transformation parameter $\omega^a(x)$ in the classical non-abelian Yang-Mills theory. The space-time dependent gauge transformation parameter $\omega^a(x)$ appears in the space-time dependent gauge transformation matrix $U_{ab}(x)$ in the classical Yang-Mills theory where the Yang-Mills color current density $j_\mu^a(x)$ of the quark transforms as \cite{yang}
\bea
j'^a_\mu(x) = U_{ab}(x)j_\mu^b(x)=[e^{M(x)}]_{ab}j_\mu^b(x),~~~~~~~~~~~~~M_{ab}(x)=f^{abc}\omega^c(x)
\label{jtr}
\eea
where $j_\mu^a(x)$ satisfies the equation
\bea
D^\mu[A]j_\mu^a(x)=0,~~~~~~~~~~D_\mu^{ab}[A]=\delta^{ab}\partial_\mu +gf^{acb}A_\mu^c(x).
\label{cceq}
\eea
Hence any gauge transformation parameter $\omega^a(x)=\beta^a(t)$ in the classical Yang-Mills theory has to satisfy the eqs. (\ref{jtr}) and (\ref{cceq}).

\section{Constant Color Charge ${\vec q}$ Produces Coulomb Potential }\label{cccp}

If all the color charges $q^a$ are constants then the classical non-abelian Yang-Mills color current density $j_\mu^a(x)$ satisfies the continuity equation \cite{ne,nj}
\bea
\partial^\nu j_\nu^a(x)=0
\label{ceq1}
\eea
which is similar to abelian-like theory (like U(1) gauge theory).

Similarly if all the color charges $q^a$ are constants then the classical non-abelian Yang-Mills potential (color potential) $A_\nu^b(x)$ in eq. (\ref{1cpx}) becomes \cite{nj}
\bea
A_\nu^b(x)= \frac{q^b~u_\nu(\tau_0)}{u(\tau_0)\cdot (x-X(\tau_0))},~~~~~~~x_0-X_0(\tau_0)=|{\vec x}-{\vec X}(\tau_0)|,~~~~~~~~u_\nu(\tau)=\frac{dX_\nu(\tau)}{d\tau}
\label{mp}
\eea
which is the abelian-like potential (like U(1) potential). From eq. (\ref{mp}) one finds that if all the color charges $q^a$ are constants then the potential energy $V(r)$ between two static color sources separated by a distance $r$ becomes
\bea
V(r)=\frac{g^2}{r}
\label{pte}
\eea
which is the Coulomb potential energy in abelian-like theory (like U(1) gauge theory).

\section{ Time Dependent Fundamental Color Charge of Quark is Not Gauge Transformation of Constant Color Charge }\label{gtnc}

If one assumes that there is a gauge choice where the color charge $q^a$ are all constants in the classical Yang-Mills theory then one may claim that the time dependent color charge $q^a(t)$ is the gauge transformation of the constant color charge $q^a$. This, however, is not true in the classical Yang-Mills theory which we will show in this section.

Let us assume that there is a gauge choice where the color charge $q^a$ are all constants in the classical Yang-Mills theory. Now making a time dependent rotation in the adjoint representation of SU(N) we find the time dependent color charge $q'^a(t)$
\bea
q'^a(t)=R_{ab}(t)q^b=[e^{\Theta(t)}]_{ab}q^b,~~~~~~~~~~~~~\Theta_{ab}(t)=f^{abc}\theta^c(t)
\label{s3p}
\eea
in SU(3) and
\bea
q'_i(t)=R_{ij}(t)q_j=[e^{\Theta(t)}]_{ij}q_j,~~~~~~~~~~~~~\Theta_{ij}(t)=\epsilon^{ijk}\theta^k(t)
\label{s2p}
\eea
in SU(2). In terms of three time dependent Euler-angles $\theta'_1(t)$, $\theta'_2(t)$, $\theta'_3(t)$ we find that the rotation matrix $R(t)$ in eq. (\ref{s2p}) is given by
\bea
R(t)=\left(\begin{array}{ccc}
{\rm cos}\theta'_1(t),&-{\rm sin}\theta'_1(t),&0\\
{\rm sin}\theta'_1(t),&{\rm cos}\theta'_1(t),&0\\
0,&0,&1\\
\end{array}\right)\left(\begin{array}{ccc}
{\rm cos}\theta'_2(t),&0,&{\rm sin}\theta'_2(t)\\
0,&1,&0\\
-{\rm sin}\theta'_2(t),&0,&{\rm cos}\theta'_2(t)\\
\end{array}\right)\left(\begin{array}{ccc}
{\rm cos}\theta'_3(t),&-{\rm sin}\theta'_3(t),&0\\
{\rm sin}\theta'_3(t),&{\rm cos}\theta'_3(t),&0\\
0,&0,&1\\
\end{array}\right)\nonumber \\
\label{rtan}
\eea
If one takes the constant color-charge vector ${\vec q}$ in SU(2) to be
\bea
\left(\begin{array}{c}
q_1\\
q_2\\
q_3\\
\end{array}\right)= \left(\begin{array}{c}
0\\
0\\
g\\
\end{array}\right)
\label{s2b}
\eea
then one finds from eqs. (\ref{s2p}) and (\ref{rtan}) that
\bea
\left(\begin{array}{c}
q'_1(t)\\
q'_2(t)\\
q'_3(t)\\
\end{array}\right)=g\left(\begin{array}{c}
{\rm sin}\theta'_2(t)~{\rm cos}\theta'_1(t)\\
{\rm sin}\theta'_2(t)~{\rm sin}\theta'_1(t)\\
{\rm cos}\theta'_2(t)\\
\end{array}\right)
\eea
which has the similar form of the time dependent color charge $q_i(t)$ of a fermion in SU(2) Yang-Mills theory as given by eq. (5) of \cite{ne}
\bea
\left(\begin{array}{c}
q_1(t)\\
q_2(t)\\
q_3(t)\\
\end{array}\right)= g\left(\begin{array}{c}
{\rm sin}\theta_2(t)~{\rm cos}\theta_1(t)\\
{\rm sin}\theta_2(t)~{\rm sin}\theta_1(t)\\
{\rm cos}\theta_2(t)\\
\end{array}\right)
\label{qts2}
\eea
where the ranges of the time dependent angles $\theta_2(t)$ and $\theta_1(t)$ are given by
\bea
\frac{\pi}{3}\le \theta_2(t)\le \frac{2\pi}{3},~~~~~~~~~~~~~~-\pi<\theta_1(t)\le \pi.
\label{rnge}
\eea
If one makes the angles time independent in eq. (\ref{qts2}) then the constant color charge ${\vec q}$ becomes
\bea
\left(\begin{array}{c}
q_1\\
q_2\\
q_3\\
\end{array}\right)= g\left(\begin{array}{c}
{\rm sin}\theta_2~{\rm cos}\theta_1\\
{\rm sin}\theta_2~{\rm sin}\theta_1\\
{\rm cos}\theta_2\\
\end{array}\right)
\label{qts2cs}
\eea
which under a time independent rotation gives
\bea
q'_i=R_{ij}q_j=[e^{\Theta}]_{ij}q_j,~~~~~~~~~~~~~\Theta_{ij}=\epsilon^{ijk}\theta^k,~~~~~~~~~~~q'_iq'_i=q_iq_i=g^2
\label{s2cs}
\eea
which is not the same as the time dependent gauge transformation in the classical SU(2) Yang-Mills theory as the constant color charges $q^a$ in eq. (\ref{s2b}) and the constant color charges $q'^a$ in eq. (\ref{s2cs}) do not belong to classical SU(2) Yang-Mills theory (see below).

Similarly, if there is a gauge choice where the $q^a$ are all constants in the classical Yang-Mills theory then the time-independence of the Casimir invariant (eq. (9) of \cite{ne}) and the form of eq. (7) in \cite{ne} suggests that also the time dependence of the color charge $q^a(t)$ for the SU(3) case is simply a time dependent rotation in the adjoint representation of SU(3) of the time-independent color charge $q^a$. Indeed, let's start from $T^a q^a(t)$, with $T^a$ the SU(3) generators in the fundamental representation (eq. (205) in \cite{ne}). The $T^a q^a(t)$ is a Hermitian matrix which can be diagonalized by a unitary transformation $e^{iT^a\theta^a(t)}$, which is equivalent to a rotation of the $q^a(t)$ in the adjoint representation of SU(3):
\bea
e^{iT^a\theta^a(t)}T^aq^a(t) e^{-iT^a\theta^a(t)}=T^aR_{ab}(t)q^b(t)=T^a[e^{\Theta(t)}]_{ab}q^b(t),~~~~~~~~~~~~~\Theta_{ab}(t)=f^{abc}\theta^c(t).\nonumber \\
\label{gts3}
\eea
The two Casimir invariants $q^a(t) q^a(t)$ and $[d^abc q^a(t) q^b(t) q^c(t)]^2$ of SU(3) are time independent in \cite{ne}. Hence if one assumes that there exists a gauge choice where all the $q^a$ are zero except $q_3$ and $q_8$ in SU(3) then $q_3$ and $q_8$ become constants.

Hence one can argue that if there is a gauge choice where all the color charges $q^a$ are constants then the time dependent fundamental color charge $q^a(t)$ of the quark can be obtained by choosing a time dependent gauge transformation of the constant color charge $q^a$. However, as we will show below, this argument is not correct because there is no gauge choice $\theta^a(t)$ where all the color charges $q^a$ are constants in the classical non-abelian Yang-Mills theory where $\theta^a(t)$ is given by eq. (\ref{s3p}).

The best way to mathematically prove that there is no gauge choice  $\theta^a(t)$ [see eq. (\ref{s3p})] in which all the color charges $q^a$ are constants in the classical SU(3) Yang-Mills theory is to look at the covariant continuity equation of the Yang-Mills color current $j_\mu^a(x)$ as given by eq. (\ref{cceq}). As mentioned in section VII any gauge transformation parameter $\omega^a(x)=\beta^a(t)$ in the classical SU(3) Yang-Mills theory has to satisfy the eqs. (\ref{jtr}) and (\ref{cceq}). For the gauge parameter choice
\bea
\omega^a(x)=\beta^a(t)
\label{bta}
\eea
the Yang-Mills color current density transforms as [see eqs. (\ref{jtr}) and (\ref{s3})]
\bea
j'^a_\mu(x) =[e^{M(t)}]_{ab}j_\mu^b(x),~~~~~~~~~~~~~~q'^a(t)=[e^{M(t)}]_{ab}q^b(t),~~~~~~~~~~~~~M_{ab}(t)=f^{abc}\omega^c(t)
\label{jtrgt}
\eea
which satisfies the equation
\bea
D^\mu[A']j'^a_\mu(x)=D^\mu[A]j^a_\mu(x)=0,~~~~~~~~~~~~~~D_\mu^{ab}[A]=\delta^{ab}\partial_\mu +gf^{acb}A_\mu^c(x)
\label{cceqgti}
\eea
where $A'^a_\mu(x)$ is the gauge transformed classical Yang-Mills potential in SU(3). Hence the gauge transformation parameter $\omega^a(x)=\beta^a(t)$ in eq. (\ref{bta}) has to be such that the eq. (\ref{cceqgti}) is satisfied in the classical SU(3) Yang-Mills theory.

Now if we assume that there is a gauge parameter choice $\theta^a(t)$ as given by the rotation matrix in eq. (\ref{s3p}) in which all the color charges $q^a$ are constants then it does not satisfy eq. (\ref{cceqgti}) which can be shown as follows. Suppose we make a rotation with the parameter $\theta^a(t)$ as given by eq. (\ref{s3p}) on the time dependent color charge $q^a(t)$ to obtain the constant color charge $q'^a$ as follows
\bea
q'^a=[e^{\Theta(t)}]_{ab}q^b(t),~~~~~~~~~~~~~~~j'^a_\mu(x)=[e^{\Theta(t)}]_{ab}j^b_\mu(x),~~~~~~~~~~~~~\Theta_{ab}(t)=f^{abc}\theta^c(t).
\label{nbd}
\eea
The color current density $j'^a_\mu(x)$ obtained from the constant color charge $q'^a$ satisfies the continuity equation [see eq. (\ref{ceq1})]
\bea
\partial^\mu j'^a_\mu(x)=0,~~~~~~~~~~~~{\rm for~constant~color~charge}~~~~~~~~~~~~q'^a.
\label{ccch}
\eea
Using eq. (\ref{nbd}) in (\ref{ccch}) we find
\bea
0=\partial^\mu j'^a_\mu(x)\neq \partial^\mu j^a_\mu(x)
\label{ccchh}
\eea
which does not satisfy eq. (\ref{cceqgti}) in the SU(3) Yang-Mills theory.

Hence we find that there is no gauge choice $\theta^a(t)$ in the SU(3) classical Yang-Mills theory in which all the color charges $q^a$ are constants where $\theta^a(t)$ is given by eq. (\ref{s3p}). This argument is valid for classical SU(N) Yang-Mills theory.

This implies that the form of the time dependent color charge ${\vec q}(t)$ in classical SU(2) Yang-Mills theory in eq. (5) of \cite{ne} is not a time dependent gauge transformation of constant color charge ${\vec q}$ but the form of ${\vec q}(t)$ in eq. (5) of \cite{ne} is the general form of a time dependent vector in the spherical polar coordinates in three dimensions [or in the adjoint representation of SU(2)]. Similarly, the form of the time dependent color charge ${\vec q}(t)$ of the quark in classical SU(3) Yang-Mills theory in eq. (7) of \cite{ne} is not a time dependent gauge transformation of constant color charge ${\vec q}$ but the form of the color charge ${\vec q}(t)$ of the quark in eq. (5) of \cite{ne} is the form of a time dependent vector in the adjoint representation of SU(3).

Note that one may be tempted to choose the normalization $g^2C_2$ with $C_2=\frac{N^2-1}{2N}$ for classical SU(N) Yang-Mills theory instead of $g^2$ as was done in \cite{ne}. However, the normalization $q^a(t)q^a(t)=g^2C_2$ is not correct in the classical Yang-Mills theory because when the color charge $q^a$ is constant then the normalization $g^2C_2$ will not reproduce eq. (\ref{pte}) for abelian theory (Maxwell theory). Since the $q^a(t)q^a(t)=g^2$ in \cite{ne} in the classical Yang-Mills theory is taken by making the analogy with the classical Maxwell theory \cite{yang,ne,nj} the normalization $q^a(t)q^a(t)=g^2$ is correct in the classical Yang-Mills theory.

\section{Gauge Invariant Color Singlet Time Dependent Potential Energy Between Static Quarks Is Not a Gauge Artifact }

As shown in sections \ref{gtcc} and \ref{gtnc} by keeping the Casimir invariants $q^a(t)q^a(t)$ and $[d_{abc}q^a(t)q^b(t)q^c(t)]^2$ time-independent, one does not find that the time dependence of the color charge $q^a(t)$ is simply a gauge transformation of the time independent color charge $q^a$. This is because if there exists any such gauge transformation then the SU(3) Yang-Mills theory will reduce to U(1) gauge theory which is not correct because one can not reduce the SU(3) Yang-Mills theory to SU(2) Yang-Mills theory or to U(1) gauge theory by making gauge transformations. Hence starting from time dependent color charge $q^a(t)$, which can be rotated by a time dependent gauge-transformation to time dependent color charge $q'^a(t)$, one finds that the gauge invariant square of the chromo-electric field in eq. (\ref{ypotef}) is time dependent.

Note that as mentioned in section \ref{cccp} when all the color charges $q^a$ are constants then the potential becomes Coulomb potential as given by eq. (\ref{pte}) which is time independent. This implies that if there is a gauge choice where all the color charges $q^a$ are constants in the SU(3) classical Yang-Mills theory then one can make such a gauge transformation to find that the gauge invariant potential energy between two static quarks separated by a distance in the SU(3) classical Yang-Mills theory is given by eq. (\ref{pte}) which can not be correct because the eq. (\ref{pte}) having the Coulomb form is for abelian-like theory. We know that the potential energy in QCD at long distance is not of the Coulomb form due to confinement at long distance. Since the potential energy in QCD at long distance and the potential energy in the classical Yang-Mills theory are same, see section III of \cite{nksum}, one finds that the potential energy in the classical SU(3) Yang-Mills theory is not of the Coulomb form. This means there is no gauge choice where all the color charges $q^a$ are constants in the SU(3) classical Yang-Mills theory. Hence we find that the gauge invariant color singlet time dependent potential energy between static quarks in the classical Yang-Mills theory in eq. (\ref{ypotef}) is not a gauge artifact.
\section{Conclusions}
Lattice QCD predicts that the potential energy between static quarks depends on the separation between quarks but is independent of time \cite{latall}. However, in this paper we have shown that the gauge invariant color singlet potential energy between static quarks in the classical Yang-Mills theory depends on time even if the quarks are at rest. This is a consequence of the time dependent fundamental color charge $q^a(t)$ of the quark in the classical Yang-Mills theory. We have found that the gauge invariant color singlet time dependent potential energy between static quarks does not violate the conservation of energy in the Yang-Mills theory.

\end{document}